\documentstyle[12pt]{article}
\begin{document}
\begin{center}
{\bf JET PHYSICS AT LEP AND QCD}

\bigskip

I.M. Dremin

{\it Lebedev Physical Institute, Moscow 119991, Russia}

\end{center}

\begin{abstract}
With the advent of LEP we discovered jets and at its last years we 
learned a lot
about intermediate vector bosons. All that strongly supports our 
belief in QCD.
Jet physics is briefly described in this talk. Experimental results are 
compared
with QCD predictions. It is shown that the perturbative QCD has 
been
able not only to {\it describe} the existing data but also to {\it predict}
many bright phenomena.
\end{abstract}

The main process of $e^+e^-$-annihilation at LEP looks like two jets 
moving in
the opposite directions. These jets are considered in QCD as 
(initiated by
quarks and collimated) cascades of consecutive emissions of 
partons each of
which produces observed hadrons due to soft confinement.\\

{\bf 1. Early days.}

Jets were discovered in 1975. Their angular collimation was 
demonstrated in
studies of such kinematical properties as sphericity, spherocity, 
thrust etc.
They show that the transverse momenta are small compared to the 
total momenta
if the proper coordinate axes are chosen. The collimation increases 
with
energy increase.

The jet axes were defined from these characteristics. It has been 
shown that
the angular ($\theta$) distribution of jet axes in $e^+e^-$-annihilation 
follows
the dependence $\propto (1+\cos ^2\theta )$ expected for spin 1/2 
objects. 

The quark origin of jets was proven in studies of the ratio of
$e^+e^-$-annihilation cross sections to hadrons and to $\mu^+\mu^-
$-pairs.
QCD predicts that this ratio should be equal to the sum of squared 
charges
of the objects initiating jets. In experiment, this ratio equals just its 
value
for quarks and increases with energy at the thresholds for heavier 
quarks.  

Thus, these early findings assured us that QCD is on the right way. 
Later,
numerous data supported this conclusion as is shown in what 
follows in brief.
To shorten the presentation, only some of most impressive results 
have been
chosen and no Figures are presented. More complete list with a 
detailed survey
and Figures demonstrating comparison with experiment can be 
found in the books
\cite{1, dkmt} and in recent review papers \cite{dre1, koch, dgar, 
kowo, dre2}.\\

{\bf 2. Theory.}

According to QCD, the primary quarks emit gluons which, in their 
turn, can emit
$e^+e^-$-pairs and gluons. Thus the branching process of jet 
evolution appears.
The gluons with high enough transverse momenta can create gluon 
jets. QCD
pretends to describe jets of both quark and gluon origin. Many 
hadrons are
created when partons become confined.

Analytical QCD pretends to start with asymptotical values and 
proceed
to lower energies accounting for conservation laws, higher order 
perturbative
and non-perturbative effects. The perturbative approach is justified 
at high
transferred momenta due to the asymptotic freedom property of 
QCD which states 
that the coupling strength becomes smaller with increase of 
transferred
momenta. The perturbative evolution is terminated at some
low scale $Q_0\sim 1$GeV for transverse momenta or virtualities of 
partons.
Here, the transition between partonic degrees of freedom at short 
distances to 
hadronic degrees of freedom at long distances (i.e., from weak to 
strong 
coupling) begins. Every experiment encodes this transition. To deal 
with it in
practice, the local parton-hadron duality (LPHD) is assumed which
declares that the distributions at the parton level describe the hadron
observables up to some constant factor. This concept originates 
from the
preconfinement property of quarks and gluons to form colorless 
clusters.
In this framework, the perturbative QCD has demonstrated its very 
high 
predictive power. It works surprisingly well when applied  for 
comparison
with experiment. 

The most general approach starts from the equation for the 
generating
functional. The generating functional contains complete information 
about any
multiparticle process. It is defined as
\begin{equation}
G(\{u\}, \kappa _0)=\sum _n\int d^3k_1...d^3k_n 
u(k_1)...u(k_n)P_n(k_1,...,k_n;
\kappa _0),
\end{equation}
where $P_n(k_1,...,k_n;\kappa _0)$ is the probability density for 
exclusive
production of particles with momenta $k_1,...,k_n$ at the initial 
virtuality
(energy) $\kappa _0$, and $u(k)$ is an auxiliary function. For 
$u(k)=$const,
one gets the generating function of the multiplicity distribution
$P_n(\kappa _0)$. The variations of $G(\{u\})$ over $u(k)$ (or 
differentials 
for constant $u$) provide any inclusive distributions and correlations 
of 
arbitrary order, i.e. complete information about the process. The 
general
structure of the equation for the generating functional describing the 
jet
evolution for a single species partons can be written symbolically as
\begin{equation}
G'\sim \int \alpha _SK[G\otimes G-G]d\Omega.
\end{equation}
It shows that the evolution of $G$ indicated by its variation 
(derivative)
$G'$ is determined by the cascade process of the production of two 
partons by a
highly virtual time-like parton (the term $G\otimes G$) and by the 
escape of a
single parton ($G$) from a given phase space region $d\Omega $. 
The weights
are determined by the coupling strength $\alpha _S$ and the splitting 
function
$K$ defined by the interaction Lagrangian. The integral runs over all 
internal
variables, and the symbol $\otimes $ shows that the two partons 
share the
momentum of their parent. This is a non-linear integrodifferential 
probabilistic
equation with shifted arguments in the $G\otimes G$ term under the 
integral sign.

For quark and gluon jets, one writes down the system of two coupled 
equations.
Their solutions give all characteristics of quark and gluon jets and 
allow for
the comparison with experiment to be done. Let us write them down 
explicitly
for the generating functions now.
\begin{eqnarray}
&G_{G}^{\prime }&= \int_{0}^{1}dxK_{G}^{G}(x)\gamma 
_{0}^{2}[G_{G}(y+\ln x)G_{G}
(y+\ln (1-x)) - G_{G}(y)] \nonumber \\ 
&+&n_{f}\int _{0}^{1}dxK_{G}^{F}(x)\gamma _{0}^{2}
[G_{F}(y+\ln x)G_{F}(y+\ln (1-x)) - G_{G}(y)] ,   \label{50}
\end{eqnarray}
\begin{equation}
G_{F}^{\prime } = \int _{0}^{1}dxK_{F}^{G}(x)\gamma 
_{0}^{2}[G_{G}(y+\ln x)
G_{F}(y+\ln (1-x)) - G_{F}(y)] ,                                   \label{51}
\end{equation}
where $G^{\prime }(y)=dG/dy ,$
$y=\ln (p\Theta /Q_0 )=\ln (2Q/Q_{0}), p$ is the initial momentum, 
$\Theta $ 
is the angle of the divergence of the jet (jet opening angle), assumed 
here to be 
small, $Q$ is the jet virtuality,  $Q_{0}=$ const , 
$ n_f$ is the number of active flavors,
\begin{equation}
\gamma _{0}^{2} =\frac {2N_{c}\alpha _S}{\pi } ,                \label{52}
\end{equation}
the running coupling constant in the one-loop approximation is
\begin{equation}
\alpha _{S}(y)=\frac {6\pi }{(11N_{c}-2n_f)y} ,
\end{equation}
the labels $G$ and $F$ correspond to gluons and quarks,
and the kernels of the equations are
\begin{equation}
K_{G}^{G}(x) = \frac {1}{x} - (1-x)[2-x(1-x)] ,    \label{53}
\end{equation}
\begin{equation}
K_{G}^{F}(x) = \frac {1}{4N_c}[x^{2}+(1-x)^{2}] ,  \label{54}
\end{equation}
\begin{equation}
K_{F}^{G}(x) = \frac {C_F}{N_c}\left[ \frac {1}{x}-1+\frac {x}{2}\right] ,   
\label{55}
\end{equation}
where $N_c$=3 is the number of colours, and $C_{F}=(N_{c}^{2}-
1)/2N_{c}
=4/3$ in QCD. The variable $u$ has been omitted in the generating 
functions.

Let us note that these equations can be exactly solved \cite{21}
if the coupling strength is assumed fixed, i.e. independent of $y$. 
For the
running coupling strength, the Taylor series expansion can be used 
\cite{13}
to get the modified perturbative expansion of physically measurable 
quantities.
The asymptotical results
are obtained in the so-called double-logarithmic (DLA) or leading 
order (LO)
approximation when the terms $(\alpha _S\ln ^2s)^n$ are summed. 
Here $s$ is 
the cms energy squared. The emitted gluons are assumed so soft 
that the
energy-momentum conservation is neglected. The corrections 
accounting for
conservation laws in the $G\otimes G$ term and in limits of the 
integration
as well as the higher order terms in the weight $\alpha _SK$ (in 
particular,
the non-singular terms of the kernels $K$) appear first in the
next-to-leading (NLO or MLLA - modified leading logarithm 
approximation) and 
then in 
higher (2NLO,...) orders. Formally, these equations have been 
proven only for
the next-to-leading (NLO) order of the perturbative QCD. However, 
one can try
to consider them as kinetic equations in higher orders and/or 
generalize them
including the abovementioned effects in a more rigorous way than it 
is usually
implied.\\

{\bf 3. QCD predictions and their comparison with experiment.}

The theoretical results have been successfully compared with 
available
experimental data. The main bulk of the data is provided by
$e^{+}e^{-}$-processes at $Z^0$ energy.

{\it The energy dependence of mean multiplicity.}

The equations for the average multiplicities in jets are obtained from
the system of equations (\ref{50}), (\ref{51}) by expanding the 
generating
functions in $u-1$ and keeping the terms
with $q$=0 and 1 according to the definition
\begin{equation}
\frac{dG}{du}\vline _{u=1}=\sum nP_n=\langle n\rangle.
\end{equation}
From their solutions one learns about the energy evolution of the 
ratio of
multiplicities in quark and gluon jets $r$ and of the QCD anomalous 
dimension
$\gamma $
(the slope of the logarithm of average multiplicity in a gluon jet) 
defined as
\begin{equation}
r=\frac {\langle n_G\rangle }{\langle n_F\rangle }\; ,\;\;\;\;\; \;\;\;
\gamma =\frac {\langle n_G\rangle ^{'}}{\langle n_G\rangle }
=(\ln \langle n_G\rangle )^{'}\; .  \label{def}
\end{equation}
They have been represented by the perturbative expansion at large 
$y$ as
\begin{equation}
\gamma = \gamma _{0}(1-a_{1}\gamma _{0}-a_{2}\gamma _{0}^{2}-
a_3\gamma _0^3)+O(\gamma _{0}^{5})
 , \label{X}
\end{equation}
\begin{equation}
r = r_0 (1-r_{1}\gamma _{0}-r_{2}\gamma _{0}^{2}-r_3\gamma 
_0^3)+O(\gamma _{0}^{4})
.  \label{Y}
\end{equation}
Using the Taylor series expansion of $\langle n\rangle $ at large $y$ 
in
the corresponding equations with (\ref{X}), (\ref{Y}) one gets the
coefficients $a_i,\, r_i$.

One of the most spectacular predictions of QCD states that in the 
leading order
approximation, where $\gamma =\gamma _0$, average multiplicities 
should
increase with energy \cite{mu1, 34, dfkh, bcmm} like
$\exp [c\sqrt {\ln s}]$, i.e., in between the power-like and logarithmic
dependences predicted by hydrodynamical and multiperipheral 
models.
Next-to-leading order results account for the term with $a_1$ in Eqn. 
(\ref{X})
\cite{web1, dktr, cdfw} and contribute the logarithmically decreasing
factor to this behavior whereas the higher order terms do not 
practically
change this dependence \cite{dg, cdnt8}. The fitted parameters in 
the final
expression are an
overall constant normalization factor which is defined by confinement 
and a
scale parameter $Q_0$. The $e^{+}e^{-}$-data are well fitted by 
such an
expression. Let us note here that the expansion parameter
$\gamma $ is rather large at present energies being about 0.4 - 0.5.

Let us stress here that the perturbative expansion in Eq. (\ref{X}) 
leads to
the modification of the perturbative expansion for $\langle n\rangle $. 
Since $\gamma $ exponentiates in $\langle n\rangle $, the so called 
modified
perturbative expansion shows up in $\langle n\rangle $. 

{\it Difference between quark and gluon jets}

The system of two equations for quark and gluon jets predicts that
asymptotically the energy dependence of mean multiplicities in them 
should
be identical. However, normalization differs, and gluon jets are more 
"active"
so that the ratio
$r=\langle n_G\rangle /\langle n_F\rangle $ of average multiplicities 
in
gluon and quark jets should tend at high energies \cite{brgu}
to the ratio of Casimir operators $C_A/C_F=9/4$. Once again,
this prediction shows how far are we now from the true asymptotics 
because
in experiment this ratio is about 1.5 at $Z^0$ energy and even 
smaller at
lower energies. The higher order terms \cite{43, web1, 
cdnt8}(calculated now
up to 3NLO) improve the agreement and approach the experimental 
value with an
accuracy about 15$\%$. The higher order terms change slightly also 
the energy
behavior of quark jets compared to gluon jets as observed in 
experiment. However,
the simultaneous fit of quark and gluon jets with the same set of 
fitted
parameters is still not very accurate. This failure is again due to
insufficiently precise description of the ratio $r$.

Let us stress here that LO and NLO terms in energy dependence of 
mean
multiplicities cancel in the ratio $r$. Thus, this is the most sensitive
measure of higher order corrections which pushes us to work at the 
limits
of our knowledge. Moreover, due to this cancellation the $\gamma 
_0^3$-terms
in $r$-expansion correspond to $\gamma _0^4$-terms in expansion 
of $\gamma $
itself, i.e. to 4NLO and not to 3NLO-terms there. And they have been 
calculated
analytically \cite{cdnt8}. Therefore the slight disagreement with 
experiment
should not surprise us very much when we work at (and, may be, 
even outside!)
the limits of applicability of the whole approach.

Moreover, QCD predicted \cite{cdnt8} that the ratio of multiplicities 
$r$ should
be smaller than the ratio of their slopes (first derivatives) which, in 
turn,
is smaller than the ratio of their curvatures (second derivatives), and 
all of
them are smaller than 2.25 and tend to this limit in asymptotics. This 
has
been confirmed by experiment as well. 

{\it Oscillations of cumulant moments.} 

The shape of the multiplicity distribution can be described by its 
higher
moments related to the width, the skewness, the kurtosis etc. The 
$q$-th
derivative of the generating function corresponds to the factorial 
moment
$F_q$, and the derivative of its logarithm defines the so-called 
cumulant
moment $K_q$. The latter ones describe the genuine (irreducible) 
correlations
in the system (it reminds the connected Feynman graphs).
\begin{equation}
F_{q} = \frac {\sum_{n} P_{n}n(n-1)...(n-q+1)}{(\sum_{n} P_{n}n)^{q}} 
=
\frac {1}{\langle n \rangle ^{q}}\cdot \frac {d^{q}G(z)}{du^{q}}\vline 
_{u=1}, 
\label{4}
\end{equation}
\begin{equation}
K_{q} = \frac {1}{\langle n \rangle ^{q}}\cdot \frac {d^{q}\ln 
G(z)}{du^{q}}
\vline _{u=1}. \label{5}
\end{equation}
These moments are not independent. They are connected by 
definite relations
that can easily be derived from their definitions in terms of the 
generating
function. In that sense, cumulants and factorial moments are equally 
suitable.

Solving the Eqns. (\ref{50}), (\ref{51}), one gets quite naturally the
predictions \cite{13, 21, 41} for the behavior of the ratio 
$H_q=K_q/F_q$.
At asymptotically high energies, this ratio is predicted to behave as 
$q^{-2}$.
However, the asymptotics is very far from our realm.
At present energies, according to QCD, this ratio should reveal the 
minimum
at $q\approx 5$ and subsequent oscillations. This astonishing 
qualitative
prediction has been confirmed in experiment (for the first time in Ref.
\cite{dabg}). Moreover, the oscillations of the moments with their 
rank have
been observed. The quantitative analytical estimates are not enough 
accurate but
the numerical computer solution \cite{lo2} reproduces oscillations 
quite well.
These new laws differ from all previously attempted distributions of 
the
probability theory. 

{\it The hump-backed plateau.}

Dealing with inclusive distributions, one should solve the equations 
for the
generating functional. It has been done up to NLO approximation.
As predicted by QCD, the momentum (rapidity $y$) spectra of 
particles inside
jets should have the shape of the hump-backed plateau \cite{adk1, 
dfkh, bcmm, adkt}.
This striking prediction of the perturbative QCD differs from the 
previously
popular flat plateau advocated by Feynman. It has been found in 
experiment.
The depletion between the two humps is due to angular ordering and 
color
coherence in QCD. The humps are of the approximately Gaussian 
shape near
their maxima if the variable
$\xi=-\ln x; \; x=p/E_j$
is used. Here $p$ is the particle momentum, $E_j$ is the jet energy. 
This
prediction was first obtained in the LO QCD, and more accurate 
expressions were
derived in NLO \cite{fweb}. Moments of the distributions up to the 
fourth rank
have been calculated. The drop of the spectrum towards small 
momenta becomes more
noticeable in this variable. The comparison with experimental data at 
different
energies has revealed good agreement both on the shape of the 
spectrum and on
the energy dependence of its peak position and width.

{\it Difference between heavy- and light-quark jets.}

Another spectacular prediction of QCD is the difference between the 
spectra
and multiplicities in jets initiated by heavy and light quarks. 
Qualitatively,
it corresponds to the difference in bremsstrahlung by muons and 
electrons 
where the photon emission at small angles is strongly suppressed 
for muons
because of the large mass in the muon propagator. Therefore, the 
intensity of
the radiation is lower in the ratio of masses squared. The coherence 
of soft
gluons also plays an important role in QCD. For heavy quarks the 
accompanying
radiation of gluons should be stronger depleted in the forward 
direction
(dead-cone or ring-like emission). It was predicted \cite{83, sdkk} 
that it should
result in the energy-independent difference of companion mean 
multiplicities
for heavy- and light-quark jets of equal energy. The naive model of 
energy
rescaling \cite{85, 86, 87} predicts the decreasing difference. The 
experimental
data support this QCD conclusion.

{\it Color coherence in 3-jet events.} 

When three or more partons are involved in hard interaction, one 
should take 
into account color-coherence effects. Several of them have been 
observed.
In particular, the multiplicity can not
be represented simply as a sum of flows from independent partons. 
QCD predicts
that the particle flows should be enlarged in the directions of 
emission of
partons and suppressed in between them. Especially interesting is 
the prediction 
that this suppression is stronger between $q\overline q$-pair than 
between
$gq$ and $g\overline q$ in $e^{+}e^{-}\rightarrow q\overline {q}g$ 
event if all
angles between partons are large (the "string" \cite{151} or "drag" 
\cite{64}
effect). All these predictions have been confirmed by experiment. In
$q\overline qg$ events the particle population values
in the $qg$ valleys are found larger than in the $q\overline q$ valley 
by a
factor 2.23$\pm $0.37 compared to the theoretical prediction of 2.4. 
Moreover,
QCD predicts that this shape is energy-independent up to an overall
normalization factor.

Let us note that for the process $e^{+}e^{-}\rightarrow q\overline 
q\gamma $
the emission of additional photons would be suppressed both in the 
direction of
a primary photon and in the opposite one. In the case of an emitted 
gluon, we
observe the string (drag) effect of enlarged multiplicity in its direction 
and
stronger suppression in the opposite one. This suppression is 
described by the
ratio of the corresponding multiplicities in the $q\overline q$ region
which is found to be equal 0.58$\pm $0.06 in experiment whereas 
the theoretical
prediction is 0.61.

The color coherence reveals itself as inside jets as in inter-jet 
regions.
It should suppress both the total multiplicity of
$q\overline {q}g$ events and the particle yield in the transverse to 
the
$q\overline {q}g$ plane for decreasing opening angle between the 
low-energy
jets. When hard gluon becomes softer, color coherence determines, 
e.g., the
azimuthal correlations of two gluons in $q\overline qgg$ system. In 
particular,
back-to-back configuration ($\varphi \sim 180^0$) is suppressed by 
a factor
$\sim 0.785$ in experiment, 0.8 in HERWIG Monte Carlo and 0.93 in 
analytical
pQCD. In conclusion, color coherence determines topological 
dependence of jet
properties.

Some proposals have been promoted for a special two-scale 
analysis of 3-jet
events when the restriction on the transverse momentum of a gluon 
jet is 
imposed \cite{egus, egkh}. They found also support from 
experiment.

{\it Intermittency and fractality.}

The self-similar parton cascade leads to special multiparton 
correlations.
Its structure with "jets inside jets inside jets..." provoked the analogy 
with
turbulence and the ideas of intermittency \cite{66}. Such a structure 
should
result in the fractal distribution in the available phase space 
\cite{drje}.
The fractal behavior would display the linear dependence of 
logarithms of
factorial moments on the logarithmic size of phase space windows. 
The moments
are larger in smaller windows, i.e. the fluctuations increase in smaller 
bins
in a power-like manner (see the review paper \cite{14}).

In QCD, the power dependence appears for a fixed coupling regime 
\cite{21}.
The running property of the coupling strength in QCD flattens 
\cite{38, 68, 70}
this dependence at
smaller bins, i.e. the multifractal behavior takes over there. The 
slopes for
different ranks $q$ are related to the Renyi dimensions. Both the 
linear
increase at comparatively large but decreasing bins and its flattening 
for
small bins have been observed in experiment. However, only
qualitative agreement with analytical predictions can be claimed 
here. The
higher order calculations are rather complicated and mostly the 
results
of LO with some NLO corrections are yet available. In experiment, 
different
cuts have been used which hamper the direct comparison. However, 
Monte Carlo
models where these cuts can be done agree with experiment better. 
The role of
partonic and hadronization stages in this regime is still debatable.

{\it Subjet multiplicities.}

A single quark-antiquark pair is initially created in $e^{+}e^{-}$-
annihilation.
With very low angular resolution one observes two jets. A three-jet 
structure
can be observed when a gluon with large transverse momentum is 
emitted by the
quark or antiquark. However such a process is suppressed by an 
additional
factor $\alpha _S$, which is small for large transferred momenta. It 
can be
calculated perturbatively. At relatively low transferred momenta, the 
jet
evolves to angular ordered subjets ("jets inside jets...").
Different algorithms have been proposed to resolve subjets. By 
increasing the
resolution, more and more subjets are observed. For very high 
resolution, the
final hadrons are resolved. The resolution criteria are chosen to 
provide
infrared  safe results.

In particular, one can predict the asymptotical ratio of subjet 
multiplicities
in 3- and 2-jet events if one neglects soft gluon coherence:
\begin{equation}
\frac {n_3^{sj}}{n_2^{sj}}=\frac {2C_F+C_A}{2C_F}=\frac {17}{8}.
\end{equation}
Actually, the coherence reduces this value to be below 1.5 in 
experiment for
all acceptable resolution parameters. Theoretical predictions 
\cite{cdfw}
agree only qualitatively with experimental findings.

Subjet multiplicities have also been studied in separated quark and 
gluon jets.
The analytical results \cite{seym} represent the data fairly well for
large values of the subjet resolution scale $y_0$.

{\it Jet universality.}

According to QCD, jets produced in processes initiated by different 
colliding
particles ($ep, pp, AA$ etc) should be universal and depend only on 
their own
parent (gluon, light or heavy quark) if not modified by the secondary
interactions. This prediction has been confirmed by many 
experiments.\\

{\bf 4. Conclusions and outlook.}
 
A list of successful analytical and Monte-Carlo QCD predictions can 
be made
longer. In particular, much work was done on the energy 
dependence of higher
moments of multiplicity distributions, on forward-backward 
multiplicity
correlations, on Bose-Einstein correlations, on various shape 
parameters of 
jets (and, in general, on event shape distributions), on non-
perturbative 
corrections etc. QCD serves not only as a powerful tool for studies of
multiparticle production processes but as a background for new 
physics as well.

The new era will be opened with the advent of new generations
of linear colliders like TESLA.\\

{\bf Acknowledgements.}

We should praise LEP and experimentalists from ALEPH, DELPHI, 
L3 and OPAL
collaborations for numerous results which allowed to enlarge our 
knowledge
of physics of multiparticle production. I regret that the available 
space
for the written version did not allow me to refer to all these papers 
and I
had to omit Figures and cite mostly the theoretical predictions. In oral
presentation, I tried to avoid these shortcomings.

\end{document}